# Greening Internet of Things for Smart Everythings with A Green-Environment Life:  A Survey and Future Prospects

S. H. Alsamhi[1], Ou Ma[2], M. Samar Ansari[3], Qingliang Meng[4]

[1]School of Aerospace Engineering, Tsinghua University & IBB University, Ibb, Yemen

[4]School of Aerospace Engineering, Tsinghua University, China

[2]College of Engineering and Applied Science, University of Cincinnati, Cincinnati, USA, and School of Aerospace Engineering, Tsinghua University, Beijing, China

[3]Department of Electronics Engineering, AMU, Aligarh, India

salsamhi@tsinghua.edu.cn, om@tsinghua.edu.cn, samar.ansari@zhcet.ac.in mengql16@mails.tsinghua.edu.cn

**Abstract:-** Tremendous technology development in the field of Internet of Things (IoT) has changed the way we work and live. Although the numerous advantages of IoT are enriching our society, it should be reminded that the IoT also consumes energy, embraces toxic pollution and E-waste. These place new stress on the environments and smart world. In order to increase the benefits and reduce the harm of IoT, there is an increasing desire to move toward green IoT. Green IoT is seen as the future of IoT that is environmentally friendly. To achieve that, it is necessary to put a lot of measures to reduce carbon footprint, conserve fewer resources, and promote efficient techniques for energy usage. It is the reason for moving towards green IoT, where the machines, communications, sensors, clouds, and internet are alongside energy efficiency and reducing carbon emission. This paper presents a thorough survey of the current on-going research work and potential technologies of green IoT with an intention to provide some clues for future green IoT research.

## I. Introduction

The internet makes the world as a small village where things are connected to each other and with the world via global communication networks using (TCP/IP) protocol. The things include not only communication devices, but also physical objects, like cars, computer, and home appliances, which are controlled through wireless communication networks. The internet has changed drastically the way we live and interact with each other in every situation spanning from professional life to social relationships [1]. Smart connectivity of the existing networks and context-aware computation using system resources is the substantial part of the internet of things (IoT). Therefore, IoT is everything around us which should be communicated ''anytime, anywhere, any media and anything".

IoT technologies make machines smarter day by day, capable of processing data intelligently and make communication more effectively and efficiently. Furthermore, IoT is a variety of things (devices), such as radio frequency identification (RFID), sensors, actuators, mobile phones, drone, etc., to communicate with each other and work together for common goals [1, 2]. Therefore, it is going to change a broad range of real-time monitoring applications such as e-healthcare, home automation, environmental monitoring, transportation autonomy, and industry automation [3, 4]. Also, IoT is an innovation in the field of wireless communication where many intelligent agents are involved sharing information, making collaborative decisions and accomplishing tasks in an optimal manner [4]. Also, IoT is all about collecting data, using data, mutual communication among devices and with the world, as shown in Fig.1. Big data requires vast storage capacity, cloud computing and large channel bandwidth for transmission; which makes IoT ubiquitous. However, big data processing consumes high power. Numerous demands for energy will, in turn, place new stresses on the society and the environment. To fulfill the smart world development and sustainability, green IoT is introduced to reduce carbon emission and power consumption.

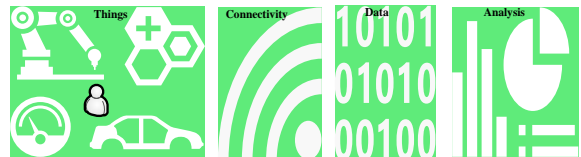

Fig.1 IoT Concept

Due to the growing awareness of environmental issues around the world, green IoT technology initiatives should be taken into consideration. Greening IoT refers to the technologies that make the IoT environmental in a friendly way by making use of facilities and storages that enabling subscribers to gather, store, access and manage various information. The enabling technologies for green IoT are called Information and Communication Technology (ICT) technologies. Green ICT technologies refer to the facilities and storages enabling subscribers to gather, store, access, and manage various information [5]. ICT technologies can cause climate change in the world [6-10] because with the growing application of ICT more and more energy has been consumed. The consideration for sustainability of ICTs has focused on data centers optimization through techniques of sharing infrastructure, which leads to increase the energy efficiency, reduce $CO_2$ emissions and e-waste of material disposals [11]. Greening ICT is enabling technologies for green IoT which includes green RFID, green wireless sensor networks (GWSN), green machine to machine (GM2M), green cloud computing (GCC), green data center (GDC) [5], green internet and green communication network as shown in Fig.2. Therefore, greening ICT technologies play an essential role to green IoT and provide many benefits to the society such as decreasing the energy used for designing, manufacturing and distributing ICT devices and equipment.

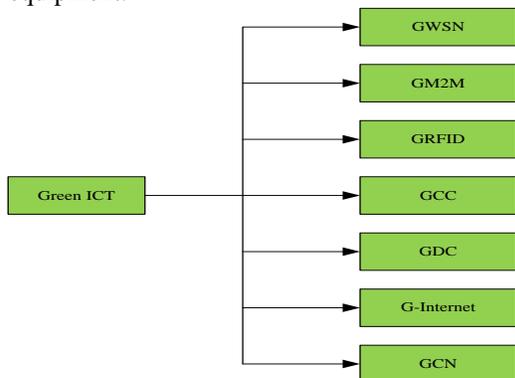

Fig.2 Green ICT technologies

Green IoT is the practice of manufacturing, designing, disposing of computers, servers, using, and associating subsystems (*i.e.,* printers, monitors, communications equipment and storage devices) efficiently and more frequently but with reduced effect on the society and the environment [12]. Going towards for greening IoT, it is looking for new resources, minimizing IoT negative impact on the health of human and disturbing the environment. The primary objective of greening IoT is to reduce $CO_2$ emission and pollution, exploit environmental conservation and minimize the costs of things operating and power consumption [13-15]. U¨ et al., [16] analyzed and provided the details about industrial emissions influence environmental change in different regions and over time. In order to make the environment healthier, reducing the energy consumption of IoT devices is required [17]. With the development of greening ICT technologies, green IoT represents a high potential to support economic growth and environmental sustainability [15]. These hot and emerging technologies make the world greener and smarter.

Surprisingly, the review has not been comprehensively carried out and the lack of a critical review in the techniques and strategies for green IoT, it is high time for a comprehensive of green IoT techniques and strategies. Therefore, this paper reviews the cores of green IoT technologies that demonstrate our work and efforts for constructing a green and smart world.

The rest of the paper is structured as follows. The overview of green IoT is presented in Section II. Sections III, IV, V, VI, VII, VIII, and IX discuss the concepts and technologies related to green RFID, GWSN, GCC, GDC, green communication networks and green internet, respectively. In Section X, applications of green IoT are listed. The future work of green IoT is discussed in section XI, and we provide concluding remarks in Section XII.

## II. Overview of Green IoT

IoT is a global, invisible, immersive, ambient communication network and computing environment built based on cameras, smart sensors, databases, software, and data centers in a world-spanning information fabric system [18]. The study in [19] adopted the idea of IoT for constructing a green campus environment aimed at energy saving. Despite prior evidence given in [19], IoT elements were discussed in [2], where the advantages of IoT architecture regarding how to create green campus by utilizing the advanced technologies smartly and efficiently were described. The authors in [20] presented specific difference technical directions towards realizing future green internet.

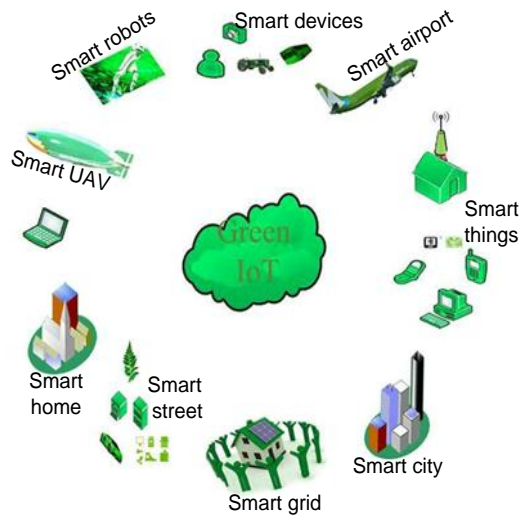

Fig.3 Green IoT

Green IoT focuses on reducing IoT energy usage, a necessity for fulfilling the smart world with the sustainability of intelligent everything and reducing $CO_2$ emissions. Green IoT consists of designing and leveraging aspects. As shown in Fig.3, design elements of green IoT refer to developing computing devices, communication protocols, energy efficiency, and networking architectures [15]. Leveraging IoT element is to reduce or eliminate emissions of $CO_2$, reduce the pollutions and enhance the energy efficiency. Despite prior evidence Uddin et al. [21] presented the techniques for enhancing the energy efficiency and reducing $CO_2$ for enabling green information technology.

Since M2M is equipped with sensors and communication add-ons, it can communicate with each other and sense the world. However, sensors will consume high power for performing the tasks. In networking, green IoT aims to identify the location of the relay and number of nodes which satisfy energy saving and budget constraints. To fulfill a smart and sustainable world, green IoT plays a significant role in deploying IoT to reduce energy consumption [5], $CO_2$ emission [22] and pollution [23-25], exploit environmental conservation [26], and minimizing power consumption [22].

Murugesan also defined the green IoT in [27] as "the study and practice of designing, using, manufacturing, and disposing of servers, computers, and associated subsystems such as monitors, storage devices, printers, and communication network systems efficiently and effectively with minimal or no impact on the environment." Green IoT has three concepts, namely, design technologies, leverage technologies and enabling technologies. Design technologies refer to the energy efficiency of devices, communications protocols, network architectures, and interconnections. Leverage technologies refer to cutting carbon emissions and enhancing the energy efficiency. Due to green ICT technologies, green IoT becomes more efficient through reducing energy, reducing hazardous emissions, reducing resources consumption and reducing pollution. Consequently, Green IoT leads to preserving natural resources, minimizing the technology impact on the environment and human health and reducing the cost significantly. Therefore, green IoT is indeed focusing on green manufacturing, green utilization, green design, and green disposal [28].

1. **Green use**: minimizing power consumption of computers and other information systems as well as using them in an environmentally sound manner.
2. **Green disposal**: refurbishing and reusing old computers and recycling unwanted computers and other electronic equipment.
3. **Green design**: designing energy efficient for green IoT sound components, computers, and servers and cooling equipment.
4. **Green manufacturing**: producing electronic components and computers and other associated subsystems with minimal or no impact on the environment.

### III. Green RFID Technology

RFID is the combined term of RF and ID where RF refers to wireless communication technology, and ID means tag identification information. It is considered as one of the promising wireless communication system used to enable IoT. Furthermore, it does not need line of sight (LoS) and can map the real world into the visual world very easy [22]. In addition, RFID is an automated data collection, and enabling objects to connect through networks which use radio waves to retrieve, identify and store data remotely. The use of electromagnetics in the radio frequency and the use of intelligent barcodes to track items in a store are referring to RFID incorporation. Storing information on things is the purpose of RFID. The classification of RFID is passive and active [22]. Passive things have not batteries on the board, and the transmission frequency is minimized. On the other hand, active RFID includes batteries that power the transmission signal.

RFID plays a vital role that helps the world to be greener by reducing the emissions of the vehicle, saving energy used and improving waste disposal, etc. Amin proposed a solution for the green RFID antennas for embedded sensors [29]. The design flexibility antenna resourcefully provides the adequate calibration of the humidity sensor following defined requirements [30]. The components of RFID are tag (carrying identification data), reader (reading/writing tag data & interfaces with a system), antenna and station (process the data) as shown in Fig.4.

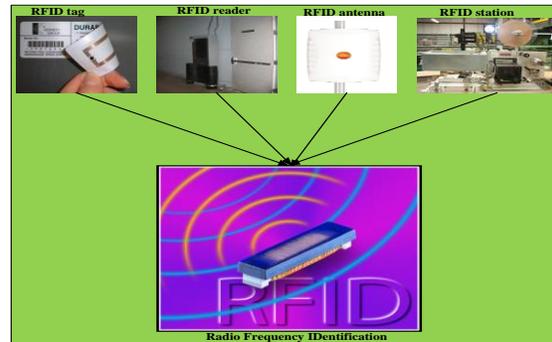

Fig.4 RFID Components

RFID usage effects positively and directly organizational agility which in turn directly and positively impacts on both logistics and operational performance [31]. Hubbard et al.[32] focused on enhancing the lifetime of unmanned aerial vehicle (UAV) battery and RFID reader detection range. The combinations of UAV and RFID are to provide additional information that can be implemented in supply chain management systems. Recharge a multi-purpose RFID tag using UAV in an environmental monitoring operation was discussed in [33]. Furthermore, Choi et al. [34] studied UAV indoor localization technique using passive UHF far-field RFID systems. UAV localization and tracking have been taken into consideration for achieving simplicity and cost efficiency. UAV is used for data collection from RFID sensors via scattered throughout the area using downloading measured data, directly approaching them and flying above them [35]. UAV and RFID sensors work together in proper way while tags can be powerful monitoring instruments. Notably, the monitoring is needed for a large area/harsh environment.

In recent years, there has been an increasing amount of researches on green RFID [5, 22, 28, 36, 37], the following suggested techniques are required for green RFID:
a) Reducing the sizes of RFID tags because recycling the tags is not easy;
b) Energy-efficient techniques and protocols should be used to avoid tag collision, tag estimation, overheating avoidance, adjusting transmission power level dynamically, etc.

There are so many applications for FRID such as transportation, production tracking, shipping, receiving, inventory control, regulatory compliance returns and recalls management. Furthermore, FRID advantages include standardized, scalable approach, reliable and cost-effective.

## IV. Green Wireless Sensor Network Technology

The combination of wireless communication and sensing has led to the wireless sensor networks (WSNs). WSNs represent the critical technology which has made IoT flourish. A sensor is a combination of an enormous number of small, low-power and low-cost electronic devices [38]. A Large number of sensors and base station (BS) nodes represent the components of WSN. Each sensor node consists of sensing, power, processing and communication unit which was discussed in [38]. Sensor nodes are being deployed around the world, measuring local and global environmental conditions such as weather, pollution, and agricultural fields and so on. Each sensor node reads from surroundings such as temperature, sound, pressure, humidity, acceleration, etc. Sensors also communicate with each other and deliver the needful sensory data to BS using ad-hoc technology. They have limited power and low processing as well as small storage capacity, while a BS node is authoritative. WSNs have various applications such as fire detection [39-41], object tracking [42-44], environmental monitoring [38, 45-47], evolving constraints in the military [48], control machine health monitoring, industrial process monitoring [38].

The idea of green IoT is supported by studies in [49, 50], which arise for keeping sensor nodes in sleep mode for most of their life to save energy as shown in Fig.5. WSNs can be just realized when data communication occurs at ultra-low power. Sensors can utilize energy harvested directly from the environment such as sun, vibrations, kinetic energy, temperature differentials, etc. [51-53].

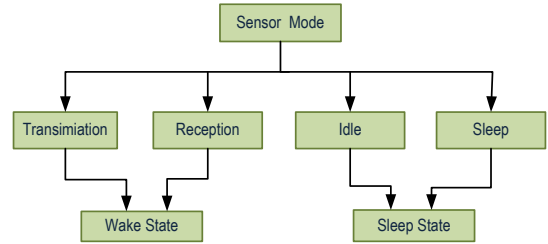

Fig.5 Sensor modes for green IoT

WSNs technology has to transmit a signal efficiently and allow going to sleep for minimal power usage. Microprocessors in sensors must also be able to wake and sleep smartly. Therefore, microprocessor trends for WSNs include reducing energy consumption; while increasing the processor speed. Therefore, green WSN is an emerging concept in which the lifespan and throughput performance are maximized while the $CO_2$ emission is pursued. The goal of WSN is supplying sufficient energy to enhance the system lifetime and contribute reliable/robust transmission without compromising the overall Quality of Service (QoS). The idea of energy efficient is supported by Mehmood et al. [54], discussed smart energy efficient of routing protocols communication for WSN concerning the design trade-offs. Similarly, Rani et al. [13] argued with details for hierarchical network design and energy efficient and flexible IoT [13]. In addition to work of [13, 54], the authors in [55] highlighted the green WSN sensors to preserve the life and routing over the WSN. Furthermore, Yaacoub et al. [56] investigated a cooperative approach for saving energy for greening WSNs. The idea of green WSN for enabling green IoT supported by a study in [57], which focused on increasing energy efficiency, extending network lifetime, reducing relay nodes and reduction in system budget. The work was implemented in four steps which were: the creation of hierarchical system frameworks and placement of sensor/actuator nodes, clustering the nodes, creation of optimization model to realize green IoT and finally the calculation of minimal energy among the nodes. The findings showed that the proposed approach was pliable, energy-saving and cost-effective when compared with the existing WSNs deployment schemes. Therefore, it is well suited for the green IoT.

The integrated and distributed clustering mechanism with real-world applications was elaborated comprehensively in [38]. The integration techniques can significantly increase the lifetime of the sensor nodes when using in environmental applications. Sensors are necessary for environmental monitoring and greenhouse control [58]. Furthermore, they should make a distributed form, spreading in the greenhouse by using distributed clustering [58]. The green designation of the actual operation and application of WSN are for minimizing the energy use. The presentation of green designation is suitable for monitoring and risk. This type of work contributed to the following [59]:
  a) Periodic data collection and notifying in contrast to sensor systems.
  b) Timestamp reconciliation and this technique dynamically supporting a growing or decreasing sensor population.
  c) Real-time visualization in a geographic context.

Mahapatra et al. [60] discussed three different schemes (i.e., wake-up radio (WUR), wireless energy harvesting (WEH) and error control coding (ECC)) to improve the performance of green WSNs while reducing the $CO_2$ emissions. However, multi-objective substructure resource allocation was proposed for green cooperative cognitive radio (CR) sensor networks [61]. On the other hand, authors in [62] proposed hybrid transmission protocol to maximize lifetime reliability guarantees for WSNs [62]. It combined the advantages of send-wait automatic repeat-request (SW-ARQ) protocol and network coding based redundant transmission (NCRT) approach. As much as possible, SW-ARQ protocol was adopted to save energy near the sink areas. In this aspect, the full use of the rest energy was made to enhance the delay and network reliability. Regarding green WSN technology, the following techniques could be adopted [5, 22, 63]:
  a. Sensor nodes should work only when necessary while spending the rest of their life in a sleep mode to save energy consumption;
  b. Energy depletion (e.g., the wireless charging, utilizing energy harvesting mechanisms which generate power from the environment such as the sun, vibrations, kinetic energy, and temperature, etc.);
  c. Optimization of radio techniques (e.g., transmission power control, cooperative communication, modulation optimization, energy-efficient CR and directional antennas);
  d. Data reduction mechanisms; and energy-efficient routing techniques.

## V. Green Cloud Computing Technology

Cloud computing (CC) is an emerging virtualization technology used across the internet. It provides unlimited computational, unlimited storage and service delivery via the internet as conceptually shown in Fig.6. CC technology is ubiquitous whereas IoT is pervasive. The combining of CC and IoT together has a broad scope of research. The primary aim of GCC is to promote the utilization of eco-friendly products which are facilely recycled and reused. Hence, Shuja et al. [64] presented an idea of green computing with a focus on information technologies. Furthermore, Baccarelli et al. [65] addressed the green economic solution in IoT over the existing Fog-supported network.

The primary purpose of GCC is to reduce the use of hazardous materials, maximize energy consumption, and enhance the recyclability of old products and wastes. Furthermore, it can be achieved by product longevity resource allocation, and paperless virtualization or proper power management. The idea is supported by a study in [28], which discuss the various technologies for GCC by reducing energy consumption. Furthermore, Sivakumar et al. [66] discussed in detail the integration of CC with IoT in various applications, architectures, protocols, service models, database technologies, sensors, and algorithms. Also, Zhu et al. [67] proposed a multi-method data delivery (MMDD) for sensor-cloud (SC) users which achieved lower cost and less delivery time. MMDD incorporates four kinds of delivery: delivery from WSN to SC users, delivery from cloud to SC users, delivery from cloudlet to SC users, and delivery from SC users to SC users.

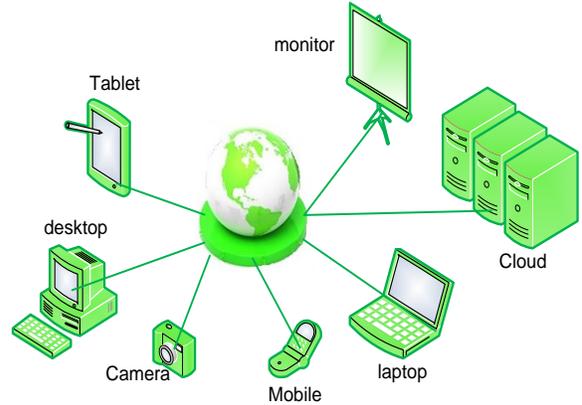
Fig.6 Green Cloud Computing

The idea of GCC is supported by applying different techniques to minimize the power requirement [68]. Authors in [68] found the important technical and analyzed the power performance of GCC and GDC. Public and private clouds were considered and included energy consumption in switching, data processing, transmission and data storage [69]. Energy consumption in transportation and circuit switching can be a significant percentage of total energy consumption in cloud computing.

Despite the numerous works in [5, 22, 70, 71] which carried out on GCC and provided potential solutions be shown as follows:
  A. Adoption of software and hardware for decreasing energy consumption.
  B. Power-saving using VM techniques (e.g., VM consolidation, VM migration, VM placement, VM allocation).
  C. Various energy-efficient resource allocation mechanisms and related tasks.
  D. Efficient methods for energy-saving systems.

E. Green CC techniques based on cloud supporting technologies (e.g., communications, networks, etc.).

## VI. Green Machine to Machine Technology

Recently, machines are increasingly becoming smarter and able to gather data without human intervention. Artificial intelligence (AI) is the drive behind the development of many recent technologies. Succeeding the idea of an intelligent machine to machine (M2M) communication is necessary to be used on a considerable scale. Machines should have good connectivity in order to enhance the modern computer machines and other electronic devices for storing large data. Then, they can share the capacity with all physical machines and any other machines around. A machine represents an object which has electrical, mechanical, environmental as well as electronic properties as shown in Fig.7. The benefit of such inbuilt radios communication is to make sure that M2M communication is safe and works efficiently for all kinds of tasks such as home, industrial, medical, as well as business processes.

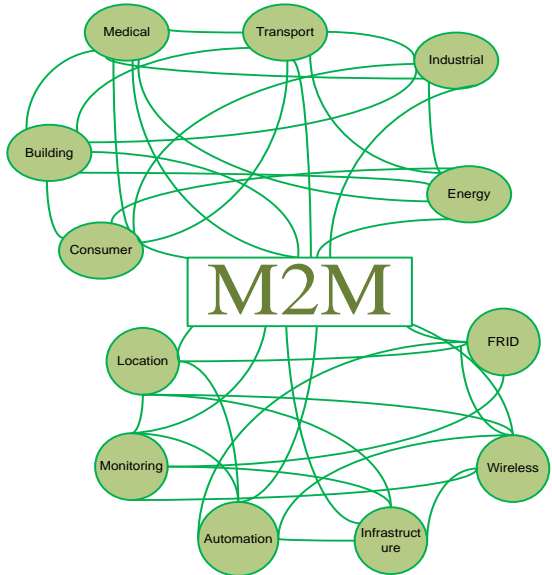

Fig.7 M2M communication

The communication between machines is described [72-74]. Therefore, a massive number of machines can communicate intelligently, share information and collaborate on decision making [75]. M2M is the advance version of IoT, where machines communicate with each other without human intervention. With the help of IoT, the billions machine can connect, recognize, communicate and respond to each other. Recent researches and projects have estimated within 5-10 years, 100 billion devices will be linked via the internet [76, 77].

Machine devices access control (MDAC) techniques used to achieve low energy consumption (EC) and also adapt to a variable distribution of MDAC [78]. A cooperative technique is proposed for improving power consumption of the cell-edge users and M2M assisted networks [79, 80]. In addition to authors in [79, 80], Bartoli et al., Datsika et al. and Dayarathna et al., discussed the necessary techniques for cooperating M2M communication network with reducing the power consumption [80-82]. The idea also supported by Himsoon et al. [83], which discussed the framework of exploiting cooperative diversity to decrease power consumption. Relay selection scheme involves determining the optimal relay node as the one that decreases the summation of transmitted power [84].

The massive M2M nodes communicate intelligently and collect data, send data to BS for deploying the M2M domain for wireless network relays.The BS further supports various M2M applications over the network in the application domain. Green M2M and the massive machines involved in M2M communications. They will consume a lot of energy, particularly in the M2M field. Niyato et al. [85] discussed the enabling technologies issues of M2M communication for the home energy management system (HEMS) in the smart grid. Several following techniques might be used to increase energy efficiency for greening IoT [5]:

a) Intelligently adjust power transmission;
b) Efficient communication protocols required for distributing the computing techniques;
c) Activity scheduling of nodes used to switch some nodes to sleeping mode while keeping the functionality of the original network;
d) Energy-saving mechanisms;
e) Employ energy harvesting and the benefits of CR.

CR is a combination of electronic network and a computer network. It is used to form an intelligent M2M communication between CR-based smart meters to remote-area power management (RAPM). The reason behind the combination is to maximize the power efficiency of electricity distribution and the spectrum efficiency [86]. Furthermore, Vo et al. [87] discussed the converged network architecture based on the flexible, high-capacity and cost-effective 4G long-term evolution (LTE) technology, which supports M2M connectivity in an end-to-end (E2E) fashion.

To decrease the energy usage, Sun et al. [88] applied clustering technique for the energy consumption (EC) of the M2M based on the random access (RA) procedure. It is carried out the overload protection and allocation of resources. The idea supported in [89], which discuss the energy efficiency implications of different emerging M2M communication standardization activities based on global standards developing organizations (GSDOs). Furthermore, using LTE technology for M2M can offload their data to neighboring helpers using the D2D communication as shown in fig.8. Fig.8 shows different machines (i.e., Drone, car, bus, plane, industries, etc.) are connected with each other via the internet. Meanwhile, relay access barring algorithm is used to enhance the performance and RA resource separation mechanism [90]. The power consumption of the M2M cooperative relay decreased, and some congestion from other M2M devices prevented. Network controlled side link communication scheme is used to enable cellular network with better support for massive machine type communication (MMTC) services [91]. Furthermore, Hussain et al. [92] focused mainly on enhancing the QoS by maximizing the admitted number of the device in the network.

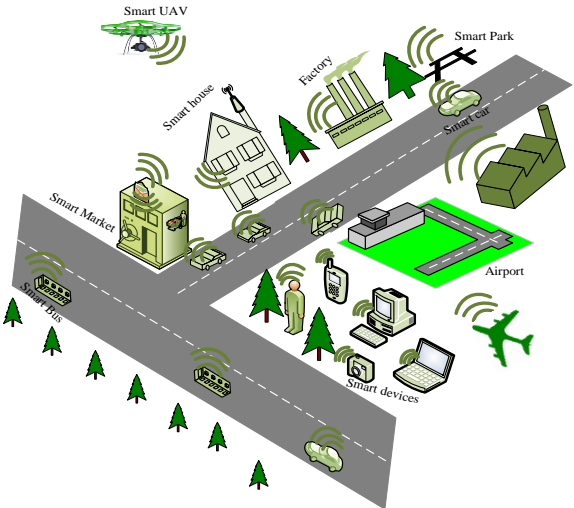

Fig.8 M2M and IoT environment

The study in [75] offers several schemes to achieve green, security and reliability in M2M connections using employing efficient activity scheduling techniques. Therefore, the prosperous of M2M communications still hinges on totally managing and understanding the current challenges: green reliability, energy efficiency, and security. Adaptive energy-harvesting MAC protocol for large M2M wireless networks improved protocol to achieve high throughput and a low transmission delay [93]. However, Abbas et al. [94] summarized studies that assessed the various IoT devices such as collision/congestion avoidance schemes, sleep and wake mechanisms to manage a device, the intelligent of select heterogeneous radio interfaces, and so on.

## VII. Green Data Center Technology

Green Data Center (GDC) is a new technology and a repository for data storage, data management, and data dissemination. These data are created by users, systems, things, etc. Dealing with different data and applications, data center (DC) consumes massive amounts of energy with high operational costs and significant $CO_2$ Footprints. Furthermore, generation of big data is rising by various ubiquitous things such as mobile devices, sensors, etc. On the way of the smart world, the energy efficiency for DC becomes more pressing [5]. Authors in [82] discussed many techniques which enhanced the energy consumption and prediction for DCs and their components [82]. In addition to work of authors [82], authors in [95, 96] presented the optimization method for the energy efficiency of DC with supporting QoS. Furthermore, GDCs are providing data services for cloud-assisted mobile ad-hoc networks (MANET) in 5G [97].

Advanced technologies are used for minimizing the building paints and carpets, low emission building materials, sustainable landscaping, using alternative energy (i.e., heat pumps, photovoltaic, and evaporative cooling). The study in [98] provides an effective method to reduce the power consumption without degrading the cooling efficiency of DCs for greening IoT. Energy saving mechanism in cloud data servers is decreasing routing and searching transactions. Peoples et al. [99] explored the mechanisms integrated effectiveness into the energy efficient context-aware broker (e-CAB) framework to manage next-generation DCs. However, the study in [100] offers a GDC of air conditioning helped by cloud techniques which consist of two subsystems: DC of air conditioning system, cloud management platform. The DC of air conditioning system includes environmental monitoring, air conditioning, communication, temperature control and ventilation; while cloud platform provides data storage, big data analysis and prediction, and up-layer application.

Ant colony system (ACS) based virtual machine (VM) is used for reducing the power consumption of DCs while preserving QoS requirements [101, 102]. The useful of using ACS was to find a near-optimal solution. Furthermore, dynamic VM is considered to reduce the energy consumption of cloud DC while maintaining the desired QoS [103]. Thus, each device is shared by many users, and VM is used to utilize those physical devices. The Mitigation of VMs for QoS constraints via minimalizing energy, bandwidth management is discussed with details in [104] and for 5G networks [105].There are several techniques used to improve energy efficiency for GDC, which can be achieved from the following aspects [5, 82, 96].

Use renewable/ green sources of energy;
a) Utilize efficient dynamic power-management technologies;
b) Design more energy-efficient hardware techniques.
c) Design novel energy-efficient data center architectures to achieve power conservation;
d) Construct efficient and accurate data center power models;
e) Draw support from communication and computing techniques.

## VIII. Green Communication and Networking

Green wireless communication plays a crucial role in green IoT. Energy efficient used by communication devices nodes should be taken into consideration. Green communications and networking refer to sustainable, energy-aware, energy-efficient, and environmentally-aware. The idea of green communication network refers to low $CO_2$ emissions, low exposure to radiation and energy efficiency. Despite prior evidence given in [106], which proposed a genetic algorithm optimization for developed the network planning, where the finding showed significantly $CO_2$ reductions cost savings and low exposure to radiation. The idea supported by a study in [61], which discuss how to maximize the data rate, minimize $Co_2$ emissions in cognitive WSNs. In addition to work of authors [61, 106], Chan et al. [107] developed a set of models for evaluating the use-phase power consumption and $CO_2$ emissions of wireless telecommunication network services. The designing of vehicular ad hoc networks (VANETs) is proposed to decrease energy consumption [108].

The feasibility of the combination of soft and green is to investigate through five interconnected areas of research (i.e., energy efficiency and spectral efficiency co-design, rethinking signaling/control, no more cells, invisible base stations, and full duplex radio) [109]. The investigation details of the energy efficiency of 5G mobile communication networks are discussed from three aspects of theory models, technology developments, and applications [110]. Furthermore, Abrol et al., [111] presented the influence, and the growing technologies need for energy efficiency in the Next Generation Networks (NGN). The need for adopting energy efficiency and $CO_2$ emission are in order to fulfill the demands for increasing the capacity, enhancing data rate and providing high QoS of the NGN. Many types of researchers have been done for saving energy by using solar and enhanced QoS [112-117]. Applying the network coding based communication and reliable storage is useful for saving energy for green IoT [118]. Stochastic geometry approach for modeling various traffic patterns can run efficiently and achieve a significant enhancement in energy efficiency while maintaining QoS requirements [119]. Utility-based adaptive duty cycle (UADC) algorithm has been proposed to reduce delay, increase energy efficiency and keep long lifetime [120]. Hypertext transfer protocol is used to enhance the lifetime and shorten the delay for providing the reliability [62]. Fig.9 shows that 5G always focuses on decreasing energy utilization and leads to green communication and healthy environments.

Nowadays, 5G may expect to impact our environment and life considerably as IoT promised to make it efficient, comfortable as shown in Fig.10. It shows the importance of 5G technology for enhancing the reliability and QoS of the communication between machines each other's and human. Furthermore, 5G technology enables to provide the large coverage connectivity, reduce the latency, saving energy and support higher data rate and system capacity. The 5G applications and its services for our society are including e-health, robotics communication, interaction human and robotics, media, transport & logistics, e-learning, e-governance, public safety, automotive and industrial systems, etc.

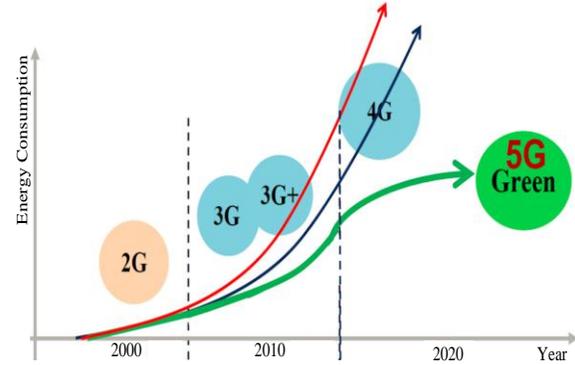

Fig.9 Developments of Energy Consumption for green Communication

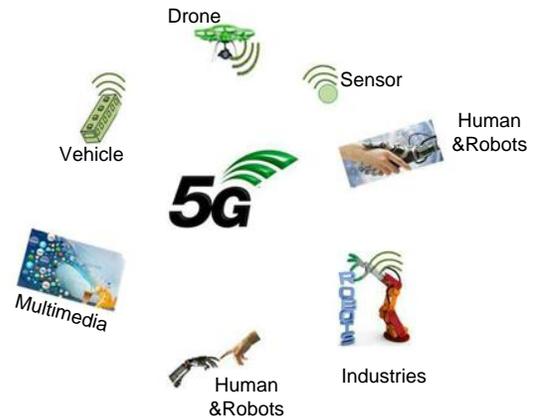

Fig.10 Expected 5G for greening IoT

## IX. Green Internet Technologies

Recently, the green internet has become the primary concern. The internet concepts and technologies contributed is used to develop a smart and green grid [121]. Reducing the power consumption efficiently is the idea of greening internet. The use of energy in the internet network equipment is unknown because of the substantial energy. The measurement of the power consumption of network equipment has explicitly been taken into account for measuring the accuracy and transparency [122]. There is a tremendous potential to decrease the internet power requirements and reduce the complexity using synchronizing the operation of scheduling traffic and routers [123]. Dynamic topology management mechanism in the green internet (GIDTMM) is built and identified the node structure and link structure for energy consumption in network device [124]. Greening internet of wired access networks (WAN) in the data network is discussed the power consumption of wired access networks estimated [125]. Furthermore, Suh at al. [126] explored an effect of the construction equipment of data network for greening the internet. The estimation power consumption and saving energy potential of data network equipment are taken into consideration.

The study in [127] designs a green internet routing technique, so the routing can lead traffic in a way that is green. Also, the idea is discussed by Yang et al. [128], which reveal the differentiated renewable and non-renewable energy for green internet routing. However, Hoque et al. [129] examined technique solutions to enhance the energy efficiency of mobile hand-held devices for the wireless multimedia streaming.

## X. Application of Green IoT

Significant changes in our environments have occurred, and some changes will occur soon because of the developments in IoT. However, the cost of the developments is potentially significant due to the increase in e-waste, hazardous emissions, and energy usage. Green IoT is estimated to make substantial changes to our future life and would lead to a green environment. In the nearest future, we will see in our daily life a lot of devices, machines, sensors, drones, and things that work and communicate with each other to accomplish their tasks intelligently for green environment. Therefore, green IoT applications have been focused on saving energy, reducing $CO_2$ emission and pollution hazardous. Not only green IoT is helping other industries reduce the greenhouse effect but also reducing the impact of IoT itself on the environment.

Green IoT benefits IoT in exploring different energy sources, eco-friendly, minimize the harm of IoT done to the environment. Thus, the numerous applications of green IoT are meaningful, economically, environmentally and social sustainability, and preserving natural resources and improving human health.

i. Smart home: A green IoT enables home equipped heating, lighting, and electronic devices to be controlled remotely by a computer/smartphone. The central mobile/computer in-house accepts voice commands. It distinguishes between residents for personalized actions and responses, Television, computer, and phone merge into one device, etc. The life cycle of green IoT should be taken into consideration that consists of the green design, green utilization, green production, and finally green disposal/recycling; to decrease the impact on the environment. In a smart home, Aslam et al.[130] formulated a model for desired QoS provisioning of heterogeneous IoT devices by using channels assigned optimization.

ii. Industrial automation: industries have been automated with machines which can do the work thoroughly automatically without or with little manual intervention based on the internet. The industrial automation based on green IoT is described briefly in [131].

iii. Smart healthcare: refers to the implementation of different biometric actuators and sensors in patients for capturing, monitoring and tracking the body of a human [132, 133]. Introducing new and advanced sensors connected to the internet for producing essential data in real-time is the IoT revolution in the healthcare industry [134, 135]. The resulting achievements of efficient health care services are enhancing the care quality, improving access to care, decreasing care costs.

iv. Smart grid: the efficiency of the smart grid is about fairness, much like the IoT. It refers to the capability of the grid dynamically adjusting and re-adjusting to deliver energy at the high quality and lowest cost optimally. A smart grid offers consumers the ability to participate in the solution. The application of communication sensor network based IoT and smart grid is discussed in [136]. Yang et al. [137] proposed a low-cost remote memory attestation for the smart grid. Furthermore, Liu et al.[138] proposed some approaches to increase data validity of IoT level data loss for smart cities. The future smart grid can measure and share energy consumption and build full energy systems [139].

v. Smart cities: represents one of the most promising and prominent IoT application [140, 141]. IoT can be characterized by efficient energy utilization to enable a sustainable smart world [142]. Hence, the machines are proposed to be equipped with additional sensory and communication add-ons to make the world smarter. Machines can sense the things surround and communicate with each other in a city. The smart city includes smart parking [143], smart light lamp [144], high-quality air, smart vehicle [145], smart traffic management [146] and so on. Maksimovic. Summarized the key to novel technology and Big data accomplishment in smart cities, where the quality of life will be improved alongside reduced pollution [147]. Smart and connected communities have evolved from the concept of smart cities [148]. The principles and applications of smart cities are discussed in [149, 150].

vi. Smart agriculture: it will enable the farmers to contend with the enormous challenges which they face. The industry should take into consideration the ways and strategies for dealing with water shortages, managing the cost, and limited land availability. Nandyala et al. [28] presented the applications of green IoT for agriculture. The combination of IoT and CC help to reduce the power consumption of the CC and IoT combination in healthcare systems and agriculture. Green IoT and green nanotechnology appear as the satisfactory solutions to create smart and sustainable agriculture and food industry [151].

## XI. Future of Green IoT

The bright future of green IoT will change our tomorrow environment to become healthier and green, very high QoS, socially and environmentally sustainable and economically also. Nowadays, the most exciting areas focus on greening things such as green communication and networking, green design and implementations, green IoT services and applications, energy saving strategies, integrated RFIDs and sensor networks, mobility and network management, the cooperation of homogeneous and heterogeneous networks, smart objects, and green localization. The following research fields have needed to be researched to develop optimal and efficient solutions for greening IoT:

1. There is a need for UAV to replace a massive number of IoT devices especially, in agriculture, traffic and monitoring, which will help to reduce power consumption and pollution. UAV is a promising technology that will lead to green IoT with low cost and high efficiency.
2. Transmission data from the sensor to the mobile cloud be more useful. Sensor-cloud is integrating the wireless sensor network and mobile cloud. It is a very hot and promised technology for greening IoT. A green social network as a service (SNaaS) may investigate for energy efficiency of the system, service, WSN and cloud management.
3. M2M communication plays a critical role to reduce energy use, hazardous emissions. Smart machines have to be smarter to enable automated systems. Machine automation delay must be minimized in case of traffic and taking necessary and immediate action.
4. Design Green IoT may be introduced from to perspectives which are achieving excellent performance and high QoS. Finding suitable techniques for enhancing QoS parameters (i.e., Bandwidth, delay, and throughput) will contribute effectively and efficiently to greening IoT.
5. While going towards greening IoT, it will be required for less energy, looking for new resources, minimizing IoT negative impact on the health of human and disturbing the environment. Then green IoT can contribute significantly to sustainable smart and green environment.
6. In order to achieve energy-balancing for supporting green communication between IoT devices, the radio frequency energy harvest should be taken into consideration.
7. More research is needed to develop the design of IoT devices which helps to reduce $CO_2$ emission and the energy usage. The most critical task for smart and green environmental life is saving energy and decreasing the $CO_2$ emission.

## XII. Conclusion

The tremendous technology development in the 21$^{st}$ century has many advantages. However, the growth of the technology demands for high energy accompanied with intention e-waste and hazardous emissions. In this paper, we survey and identify the most critical technologies used for green IoT and keeping our environment and society smarter and green. ICT revolution (*i.e.,* FRID, WSN, M2M, communication network, Internet, DC, and CC) has qualitatively augmented the capability for greening IoT. Based on the critical factors of ICT technologies, the things around us will become smarter to perform specific tasks autonomously, rendering of the new type of green communication between human and things and also among things themselves, where bandwidth utilization is maximized and hazardous emission mitigated, and power consumption is reduced optimally. Future suggestions have been touched upon for efficiently and effectively improving the green IoT based applications. This research provides effectively insight for anyone wishes to find out research in the field of green IoT. The trends and prospective future of green IoT are provided.